# Estimation of the traffic in the binary channel for data networks

Sander Stepanov

sanderstepanov@yahoo.com

**Abstract**

It is impossible to provide an effective utilization of communication networks without the analysis of the quantitative characteristics of the traffic in real time. The constant supervision of all channels of the data practically is impracticable because requires transfer of the significant additional information on a network and large resources expenses for devices of the control. Thus, the task on traffic estimation with small expenses in real time is the urgent.

**Introduction**

The dynamic management of a network is based on use on line of the information on the traffic of the users of a network. Routing, realization of payments, congestion control, realization of traffic management system - and many other functions data networks also require knowledge of the real time statistical characteristics of volume of the traffic. This work is devoted to the decision of the contradiction between necessity of the constant control of ports of the consumers of a network and dearness of its realization.

We assume that the output of the channel can be in a condition a package is transferred and in a condition the package is not transferred (the binary channel) and we know length of a package got under the control. This assumption can be used as simplification if we are interested in the traffic of various sources of the information in one channel of communication. I.e. it is possible to divide the channel on some virtual channels. And for example, if have pulled out a package we register presence of a package for the appropriate virtual channel and absence of a package for all other channels. The random processes in system of communication are assumed stationary.

The basic task is the definition of the packages transfer time as with it is possible with the greater accuracy as soon as possible at the small time of supervision of the channel. Physically it is necessary when the device of an estimation of the traffic can not constantly observe the channel, for example owing to service of several channels serially. Or when the control is carried out by means of transfer by the consumer to the device of an estimation of the traffic of the information about the condition on the channel of communication. Clearly as in this case constantly to transfer the service information is expensive (even if to transfer the information on the characteristics of a package if there was a package in a control point, and to not transfer anything if in a control point there was no package i.e. to make the decision on absence of a package by default) and consequently we are interested as small as possible to occupy the channel of communication.

The algorithm of estimation should carry out the following:

1. To predict potential accuracy of calculation of the traffic. It allows before realization of an estimation to know how many resources it is necessary to spend on estimation for probable sizes of the traffic.

2. To calculate accuracy of an estimation during calculations. It allows to operate process of an estimation in an operating time of algorithm. For example it is necessary to decide to continue supervision of the channel or to stop.

## 1 Background

We shall use model of supervision of the traffic when samples take out in random points on an axis of time on an output of a source of packages.

For convenience of the analysis, as the characteristic of a source we shall use the statistical characteristic of its ability to transfer packages. It is probability of presence of a package on an output of a source at the moment of the control, we shall designate it by a symbol U (utilization factor). Thus if we shall calculate U and after that the time t that has passed it is possible to consider that in $U*t_1$ time units the packages were transferred. It will be similarly possible to predict that during the future $t_2$ time units to be transferred packages $U*t_2$ time units.

We pay attention that instead of search of the traffic transferred during work we search for the characteristic of ability of a source to transfer packages, i.e. we assume presence of such hypothetical characteristic, that in mathematical statistics and theory of probability the widespread assumption. For example, binomial experiments.

From above-stated follows, that us should not confuse if as a result of work will is made a method of definition of the traffic and also method of calculation of accuracy of definition of the traffic. And if we shall look that will give these methods at constant supervision of a source, notwithstanding what we shall observe all time and should know now traffic precisely, but in these methods will give errors in definition U, it because in model unknown not the traffic but U.

One of the most powerful statistical methods is the method Maximum likelihood estimation (MLE). According to this method, the account of dependence of supervision is complex, therefore at first we shall create a method for independent samples, and then we shall develop a method of realization of independence. The independence of samples makes also lot of the information about the traffic in the limited quantity of samples in comparison with a case of dependent samples and consequently provides reduction of time of estimation.

Let n - quantity of points of the control, y - quantity of points in which the packages were observed. Then probability of supervision of packages in y points.

$$P_{y,n} = C_n^y U^y (1-U)^{n-y} \quad (1.1)$$

Let's find expression for U, providing the greatest probability of packages in y points

$$\hat{U} = \arg\max_{U} C_n^y U^y (1-U)^{n-y} \quad (1.2),$$

it is equivalent

$$\hat{U} = \arg\max_{U} \ln[U^y (1-U)^{n-y}] \quad (1.3).$$

Solving (1.3), we can write

$$\frac{d[y \ln \hat{U} + (n-y)\ln(1-\hat{U})]}{d\hat{U}} = 0 \quad (1.4)$$

then obtain

$$\hat{U} = y/n \quad (1.5).$$

Lets $\xi$ - casual variable equal 1 when in a point of the control there is a package and equal 0 when in a point of the control there is no package,

$$y = \sum_{i=1}^{i=n} \xi_i \qquad (1.6).$$

Mean $\hat{U}$

$$M(\hat{U}) = M(y/n) = \frac{1}{n} * M(y) = \frac{1}{n} * \sum_{i=2}^{i=n} M(\xi_i) = nU/n = U \qquad (1.7),$$

the characteristics of variance $\hat{U}$

$$VAR(\hat{U}) = VAR(y/n) = \frac{VAR(y)}{n^2} = \frac{\sum_{i=1}^{i=n} VAR(\xi_i)}{n^2} = \frac{nU(1-U)}{n^2} = U(1-U)/n \qquad (1.8)$$

$$\frac{dVAR(\hat{U})}{dU} = \frac{1-2U}{n} \qquad (1.9)$$

$$VAR(\frac{\hat{U}}{U}) = \frac{VAR(\hat{U})}{U^2} = (\frac{1}{U} - 1)/n \qquad (1.10)$$

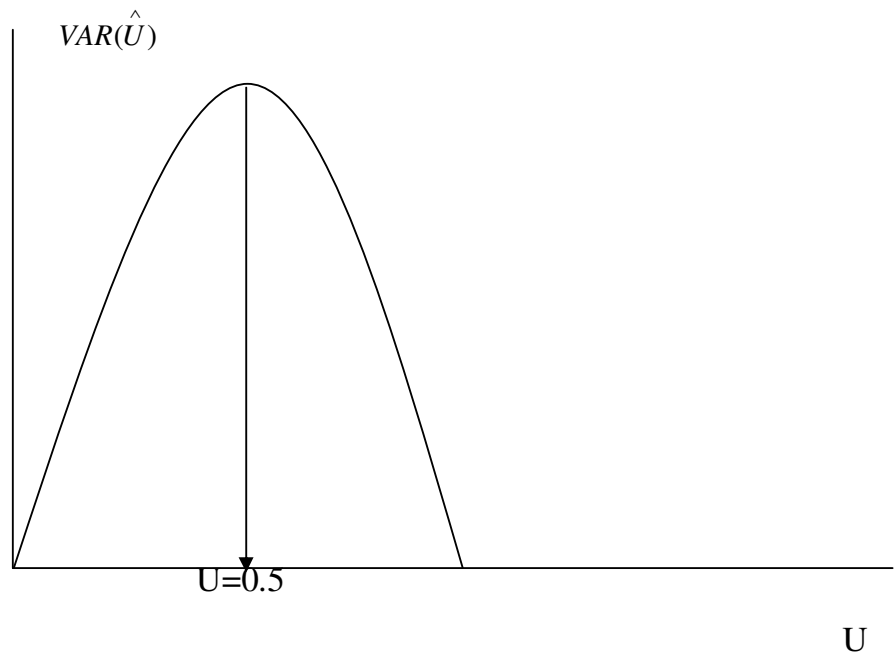

Fig. 1

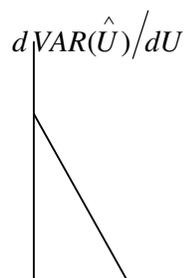

U=0. 5

U

Fig. 2

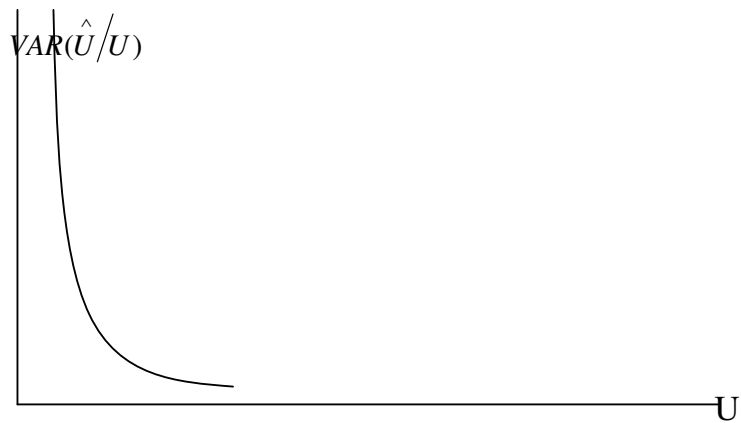

Fig. 3

As it is visible from the (1. 8), (1.9) and figures Fig.1, Fig.2 for, absolute variance small for small and large values U and to become maximal at U=0.5. However sometimes better for practice to be interested relative variance. From the formula (1.10) and figure Fig.3 follows that at small values U the mistake of calculation considerably grows.

If before extraction of samples to determine their sites by the generator of random numbers with uniform distribution, given on a prospective time interval of the analysis, then a task of definition of the traffic to become similar to a task of calculation of the certain integrals. As the calculated area there will be a time during packages. This reduction of a task of definition of the traffic to the well developed task of calculation of integrals by a method Monte Carlo can be used when we are not interested by the characteristic of productivity of a source and we agree constantly to supervise an output of a source, instead of definition of the characteristic of productivity of a source and subsequently only to check it has changed. At application of this simplification, the special attention on is necessary that the principle of extraction of samples must correspond to a method of calculation of integrals. The complete conformity would be if we could take samples not under the temporary order, as at definition of the traffic, but by way of generation of random numbers, as at calculation of integrals.

For analyze we shall designate: T- common operating time of a source; $T_p$ - common time during which the packages were transferred; $\bar{t}_p$ - mean size packet; $\bar{t}_b$ - mean interval between packets. Then probability observation packet for method Monte Carlo

$$U_{MC} = T_p/T \quad (1.12).$$

Variance estimate $T_p$

$$\hat{T}_p = T\frac{y}{n} = T\hat{U}, \quad (1.13)$$

we cane calculate using (1.8)

$$VAR(\hat{T}_p) = VAR(\hat{U}) * T^2 = T_p(T - T_p)/n \quad (1.14).$$

The analysis of equality (1.14) is done similarly to equality (1.8) and the results it is similar. In case of interest to relative variance

$$VAR(\frac{\hat{T}_p}{T_p}) = \frac{T^2}{T^2_p}VAR(\hat{U}) = (\frac{T}{T_p} - 1)/n \quad (1.15)$$

Considering that

$$T = N(\bar{t}_p + \bar{t}_b) \quad , \quad T_p = N\bar{t}_p \quad (1.16),$$

where    N - real quantity of packages.

Then we deduce

$$VAR(\frac{\hat{T}_p}{T_p}) = \bar{t}_b/(\bar{t}_p n) \quad . (1.17)$$

For $\bar{t}_b$ from an exponential distribution with parameter $\lambda$

$$VAR(\frac{\hat{T}_p}{T_p}) = 1/(\bar{t}_p n \lambda) \quad . (1.18)$$

Can be interesting estimation $\lambda \to \hat{\lambda}$, using

$$\overline{t_b} = 1/\lambda = \frac{T - T_p}{N} = \frac{T - T_p}{T_p / \overline{t_p}} \quad (1.19)$$

and also that instead of $T_p$ we can use (1.13), obtain

$$\hat{\lambda} = \frac{y/n}{t_p(1 - y/n)} \quad . \quad (1.20)$$

The accuracy of calculations U in process estimating can be supervised by a method confidence limits. Upper confidence limit $U_u$ satisfies to expression

$$\frac{1-\gamma}{2} = \sum_{i=0}^{i=y} C_n^i U_u^i (1 - U_u)^{n-i} . \quad (1.21)$$

Lower confidence limit $U_l$ satisfies to expression

$$\frac{1-\gamma}{2} = \sum_{i=y}^{i=n} C_n^i U_l^i (1 - U_l)^{n-i} \quad (1.22)$$

were - $\gamma$ reliability

$$\Pr(U_l \langle \hat{U} \langle U_u) \geq \gamma . \quad (1.23)$$

In figures 6 the possible variants of curve functions are shown. It is visible that at the decision of the equations by means of search methods of optimization at large n (figure 6a) difficulty to execute search of the decision because of sharp recession of functions on a short site. As has shown modeling a search method of "division on half" sometimes misses by sites with abrupt recession.

Sander

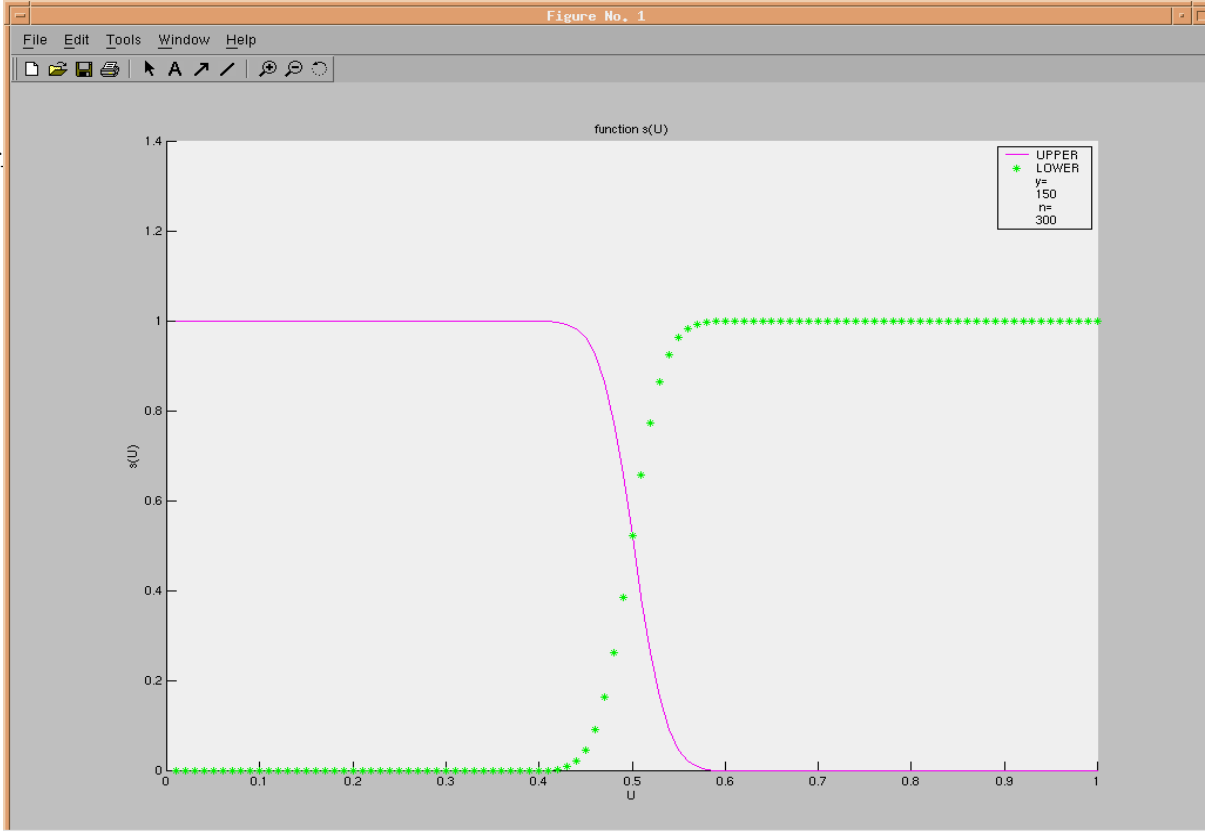

Figures 6a

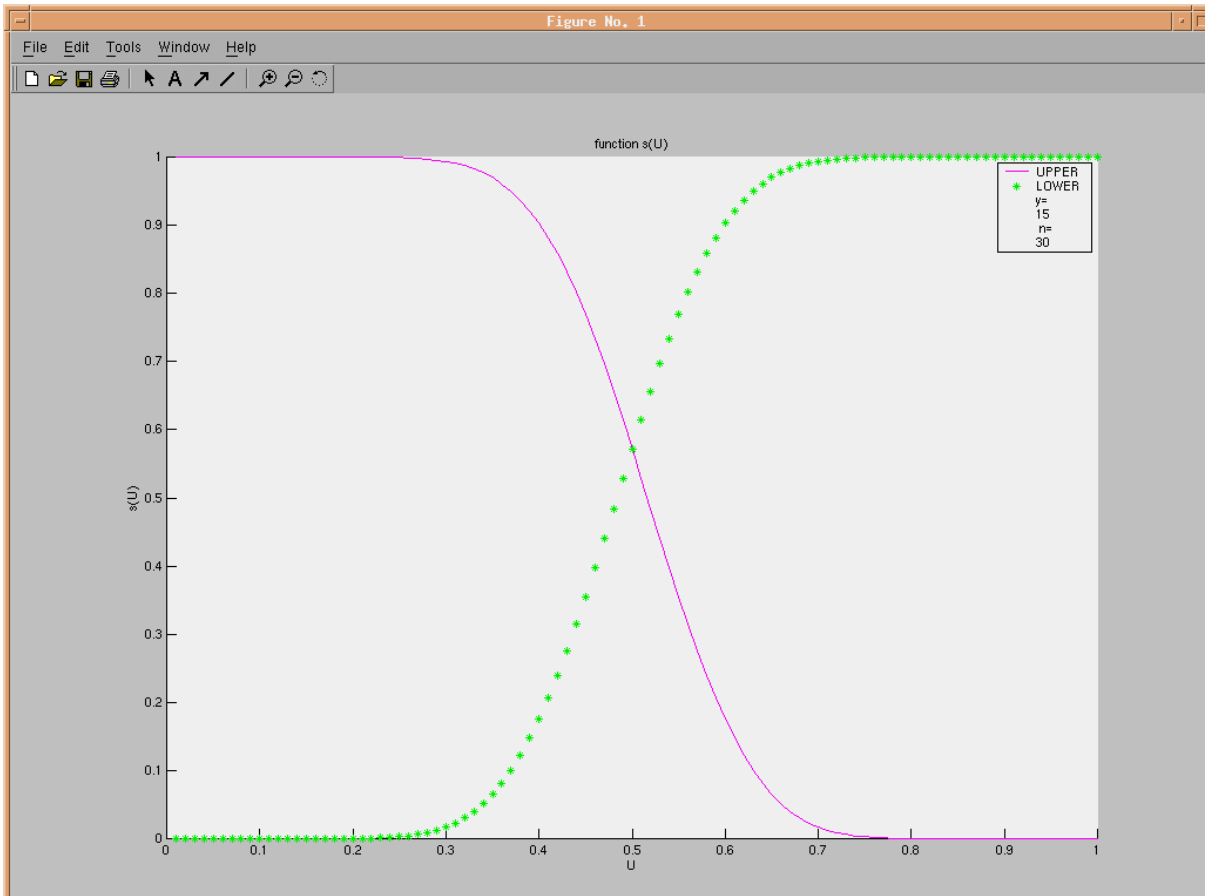

Figures 6b

Figure 6. Characteristic diagrams of functions $s(U) = \sum_{i=0}^{i=y} C_n^i U_u^{\,i} (1-U_u)^{n-i}$ and $s(U) = \sum_{i=y}^{i=n} C_n^i U_l^{\,i} (1-U_l)^{n-i}$ for a case n large (figure 6a) and case n small (figure 6b).

Thus, in the assumption of independence of samples the expression for U was is advanced. The way of development of independence of supervision is described further.

## 1. 2. Obtaining independence of samples

### 1. 2. 1. Obtaining independence by Bayes' Theorem

To obtain straight PDF of intervals between sampling points with independence between samples is complex, we shall invent therefore decision for the following experience. We make extraction of a sample in a point 1, and then in a point 2, distance between points is distributed on PDF $\varphi(t)$. It is visible that we have simplified a task, but however physical sense basically have kept. The basic researched mechanism has not changed - as well as in a real task there is a dependence between supervision in the next points, this dependence is done by size of packages and size of intervals, the relation between sizes of packages both intervals between packages and intervals between sampling points are set PDF's.

In the beginning, we shall solve a task for a case of the known statistical characteristics of a source of packages. So is considered known PDF of intervals of time between packages, PDF of lengths of packages. Let's designate them accordingly $b(\tau)$ and $f(l)$. PDF distance between points $\varphi(t)$ is unknown, by its choice we should achieve statistical independence of samples. Let's consider $\tau$, t and l as independent.

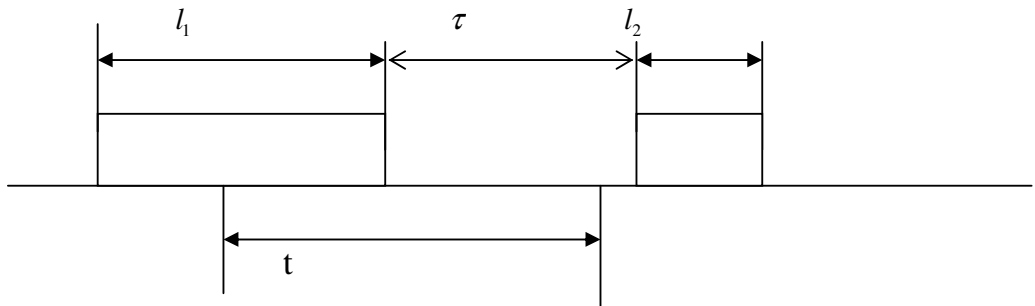

Fig. 4

In an ideal case the probability of hit of the second point a package, after the first point has got in a package, should be equal simply to probability of hit of a point in a package.

$$U = \frac{\int_0^\infty f(l)l\,dl}{\int_0^\infty f(l)l\,dl + \int_0^\infty b(\tau)\tau\,d\tau} \qquad (2.1)$$

Let's consider that the first point has got in a package, there are no reasons to believe that any place of a package more preferably for hit. Therefore, it is possible to use uniform PDF for calculation of probability of presence of the first point in any place of a package. Probability of hit of the second point a package, after the first point has got in a package is equal to the sum of not joint events: the second point has got in the same package (we shall designate its probability through $P_0$);

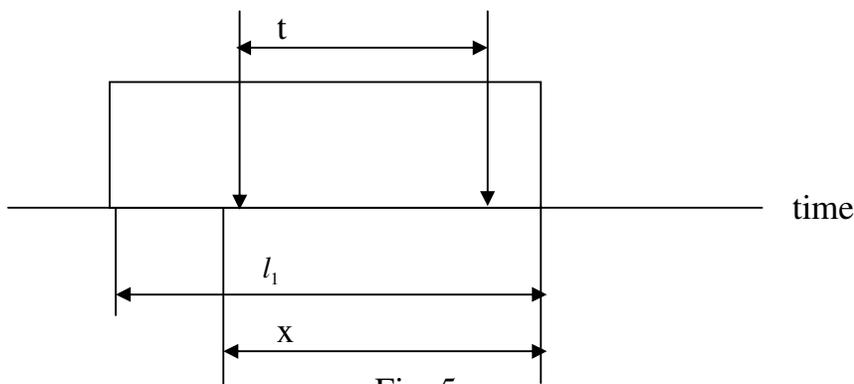

Fig. 5

$$P_0 = \int_0^\infty f(l_1) \int_0^{l_1} 1/l_1 \int_x^{l_1} \varphi(t)\,dt\,dx\,dl_1 \qquad (2.2a)$$

the second point has got in the first following package (we shall designate his(its) probability through $P_1$);

$$P_1 = \int_0^\infty f(l_1) \int_0^{l_1} 1/l_1 \int_0^\infty b(\tau) \int_0^\infty f(l_2) \int_{x+\tau}^{l_2+x+\tau} \varphi(t)\,dt\,dl_2\,d\tau\,dx\,dl_1 \qquad (2.2b)$$

etc..

In an ideal case of independent samples should be,

$$U = \sum_{i=1}^{i=\infty} P_i, \qquad (2.3)$$

however such decision, as it is visible, is very complex.

We simplify an ideal method of calculation consisting in use of expressions (2.2) - (2.3) by that we shall take into account cases when the previous point of the control was in the beginning of a package or in the beginning of a interval between packages. And that we shall search for minimally constant values of the long between points of supervision. It is possible to use variable distance between supervision but with a condition to not exceed distance of maintenance of independence.

Let's calculate distances between points of the control appropriate to these cases and we shall use their average. Correctness of the made assumptions again we shall check up by modeling.

Let previous point was in the beginning of a package, then the current point will get in a package if: (1) lengths of a package will appear more than distance of the control H, this probability is equal

$$\int_H^\infty f(\overline{V}^{real}, l) dl \qquad (2.4),$$

where $\overline{V}^{real}$ real value of a vector of parameters

(2) or length of the first package and interval after it will appear less than distance of the control and long of the second package large enough that in it the point of the control has got, this probability is equal

$$\int_0^H f(\overline{V}^{real}, y) \otimes b(\overline{Z}^{real}, y) \int_{H-y}^\infty f(\overline{V}^{real}, l) dl dy. \qquad (2.5)$$

where $\overline{Z}^{real}$ real value of a vector of parameters

Generalizing the formulas (2.4) - (2.5) we make general probability of hit in a package if the first point was in the beginning of a package

$$P_{all} = QF(\overline{V}^{real}, l, H) + \sum_{n=1}^{n=\infty} \int_0^H QF(\overline{V}^{real}, l, H-y) bf1(n, y) dy, \qquad (2.6)$$

were

$$QF(\overline{V}^{real},l,z) = \int_z^\infty f(\overline{V}^{real},l)dl \ ,$$

$$bf1(n,y) = \underbrace{f(\overline{V}^{real},y) \otimes b(\overline{Z}^{real},y) \otimes f(\overline{V}^{real},y) \otimes \circ \circ \circ \otimes b(\overline{Z}^{real},y)}_{2n}.$$

The general probability of hit in a package if the first point was in the beginning of an interval between packages to be deduced similarly.

Actually are not known PDF of lengths of packages and distances between packages. If they were are known it is possible would be to calculate U on equality (2.1).

The deduced above expressions it is possible to try to use as follows. To apply calculations by Bayes' theorem . In calculations to be set in the beginning by prior values f(l) and b($\tau$), then to calculate H.

It is Algorithm 1 to obtain H.

Step 1. Empirical choice: kind of function f(l) (let's designate its vector of parameters $\overline{V}$) and b($\tau$) (let's designate its vector of parameters $\overline{Z}$); ranges of definition for f(l) and b($\tau$); $m_{j,1}(\overline{V})$ and $m_{j,1}(\overline{Z})$ - lows bounds of parameter j for $\overline{V}$ and $\overline{Z}$, $m_{j,2}(\overline{V})$ and $m_{j,2}(\overline{Z})$ uppers bounds of parameter j for $\overline{V}$ and $\overline{Z}$, $m_{j,3}(\overline{V})$ and $m_{j,3}(\overline{Z})$ - quantity of values in an interval $[m_{j,2}(\overline{V})-m_{j,1}(\overline{V})]$ and $[m_{j,2}(\overline{Z})-m_{j,1}(\overline{Z})]$ of parameter j for $\overline{V}$ and $\overline{Z}$; prior probabilities for all values $P_{prior_0}(\overline{V},\overline{Z})$    $\forall \overline{V}, \forall \overline{Z}$;

Step 2.

$$(\overline{V}_i^{max},\overline{Z}_i^{max}) = \arg\max_{(V_{r_w},Z_{a_q})} P_{prior_o}(V_{r_w},Z_{a_q}) \quad \text{if } i=0 \quad (2.7)$$

$$(\overline{V}_i^{max},\overline{Z}_i^{max}) = \arg\max_{(V_{r_w},Z_{a_q})} P_{prior_i}(V_{r_w},Z_{a_q}) \quad \text{if } i \neq 0 \quad (2.8)$$

were    i – number of circle of algorithm;

$$V_{r_w} = m_{w,1}(\overline{V}) + \frac{m_{w,2}(\overline{V}) - m_{w,1}(\overline{V})}{m_{w,3}(\overline{V})} r, \ w=0,\ldots,Y; \ r=0,\ldots,m_{w,3}(\overline{V});$$

$$Z_{a_q} = m_{q,1}(\overline{Z}) + \frac{m_{q,2}(\overline{Z}) - m_{q,1}(\overline{Z})}{m_{q,3}(\overline{Z})} a, \quad q = 0,...,Q; \quad a = 0,...,m_{q,3}(\overline{Z});$$

Y – number of parameters of f(l); Q - number of parameters of b($\tau$);

### Step 3

$$\overline{H}_i = \min \ (\underset{\overline{H}}{\arg \ \min} \ |P_{ideal_i} - P_i|) \quad (2.\ 9)$$

$$\overline{H}_i^{approx} = \min \ (\underset{\overline{H}}{\arg \min} \ \left[|P_{ideal_i} - P_i| < K\right]) \quad (2.\ 10)$$

where  $P_{ideal_i}$ - ideal exact probabilities calculated by means of (2. 1) on step i;

$P_i$ - Probability being average between probability for a case when the first point in the beginning of a package and for a case when the first point in the beginning of an interval between packages on step i;

$\overline{H}_i$ - distance between points of the control

K - empirical number, it can be applied to simplification of optimization at finding of the decision (2. 12), because there is a limit of accuracy of calculation (2. 12) after which increase of accuracy is not important, i.e. the error at calculation (2. 12) should be not less errors of other steps of algorithm.

Second at the left "min" in the formulas (2. 9), (2. 10) means that it is necessary to search H making for the same probability to get in a package by the second point as probability to get in a package by one isolated point. First at the left "min" means that is necessary to search for the minimal H satisfying to the second condition. This second "min" provides reduction of common time of supervision behind the channel.

For understanding of computing complexity, the equality (2. 9) is shown in the complete form

$$\overline{H}_i = \min \left( \arg\min_H \left| \left[ \frac{\int_0^\infty f(\overline{V}_i^{max}, l) l \, dl}{\int_0^\infty f(\overline{V}_i^{max}, l) l \, dl + \int_0^\infty b(\overline{Z}_i^{max}, \tau) \tau \, d\tau} \right] - \right. \right.$$

$$-\frac{1}{2} \left[ QF(\overline{V}_i^{max}, l, H) + \sum_{n=1}^{n=\infty} \int_0^H QF(\overline{V}_i^{max}, l, H-y) bf1(n, y) dy \right] * \quad (2.11)$$

$$* \left[ \int_0^H QF(\overline{V}_i^{max}, l, H-y) b(\overline{Z}_i^{max}, y) dy + \sum_{n=1}^{n=\infty} \int_0^H QF(\overline{V}_i^{max}, l, H-y) bf2(n, y) dy \right] \left. \right| \right)$$

were

$$QF(\overline{V}_i^{max}, l, z) = \int_z^\infty f(\overline{V}_i^{max}, l) dl$$

$$bf1(n, y) = \underbrace{f(\overline{V}_i^{max}, y) \otimes b(\overline{Z}_i^{max}, y) \otimes f(\overline{V}_i^{max}, y) \otimes \bullet\bullet\bullet \otimes b(\overline{Z}_i^{max}, y)}_{2n} \text{ - Assumed probability}$$

of hit in a package under condition of hit of the first point in a beginning of a package;

$$bf2(n, y) = \underbrace{b(\overline{Z}_i^{max}, y) \otimes f(\overline{V}_i^{max}, y) \otimes b(\overline{Z}_i^{max}, y) \otimes f(\overline{V}_i^{max}, y) \otimes \bullet\bullet\bullet \otimes b(\overline{Z}_i^{max}, y)}_{2n+1} \text{ - Assumed}$$

probability of hit in a package under condition of hit of the first point in a beginning of an interval between packages.

For n > 6 it is possible to use normal approximation PDF sums of functions then

$$bf1(n, y) = nor(M_1*n + M_2*n, V_1*n + V_2*n) \quad (2.12)$$

$$bf2(n, y) = nor(M_1*n + M_2*(n+1), V_1*n + V_2*(n+1)) \quad . (2.13)$$

Where $M_1$, $M_2$ means and $V_1$, $V_2$ variances $f(\overline{V}^{real}, l)$ and $b(\overline{Z}^{real}, y)$ accordingly.

At use n>6 we lose if exist smaller H appropriate smaller n that we shall miss them. When the benefit from reduction of time of an estimation of the traffic large then is necessary to calculate H by more complex but also more effective way as is described below.

for $f(\overline{V}_i^{max}, l) = nor(l, M, V)$ , $b(\overline{Z}_i^{max}, \tau) = \lambda \, e^{-\lambda \tau}$ - exponential PDF, deduce

$$bf1_{n,e}(n, y) = nor(y, n*M, n*V) \otimes \frac{\lambda(\lambda y)^{n-1}}{(n-1)!} e^{-\lambda y} =$$

(2. 14)

$$= \int_0^y nor(z, n*M, n*V) \frac{\lambda(\lambda(y-z))^{n-1}}{(n-1)!} e^{-\lambda(y-z)} dz$$

$$bf1_{n,e}(n, y) = nor(y, n*M, n*V) \otimes \frac{\lambda(\lambda y)^n}{n!} e^{-\lambda y} =$$

(2. 15)

$$= \int_0^y nor(z, n*M, n*V) \frac{\lambda(\lambda(y-z))^n}{n!} e^{-\lambda(y-z)} dz$$

The results of calculations of probabilities of hit in a package are below shown when the previous point was in the beginning of a package s1(x) and in the beginning of an interval between packages s2(x) (Fig. 7, Fig 10) and program their executed.

The research of expressions (2. 11), (2. 14), (2. 15) shows that at calculations of enough small values n, so for example in figure the values of expression (2. 14) for n=1, 2,… 5 are shown when $\overline{t_b} > \overline{t_p}$ (M=3, $\lambda = 0.1$) and when $\overline{t_b} < \overline{t_p}$ (M=3, $\lambda = 1$). It is visible what enough to take into account insignificant quantity composed.

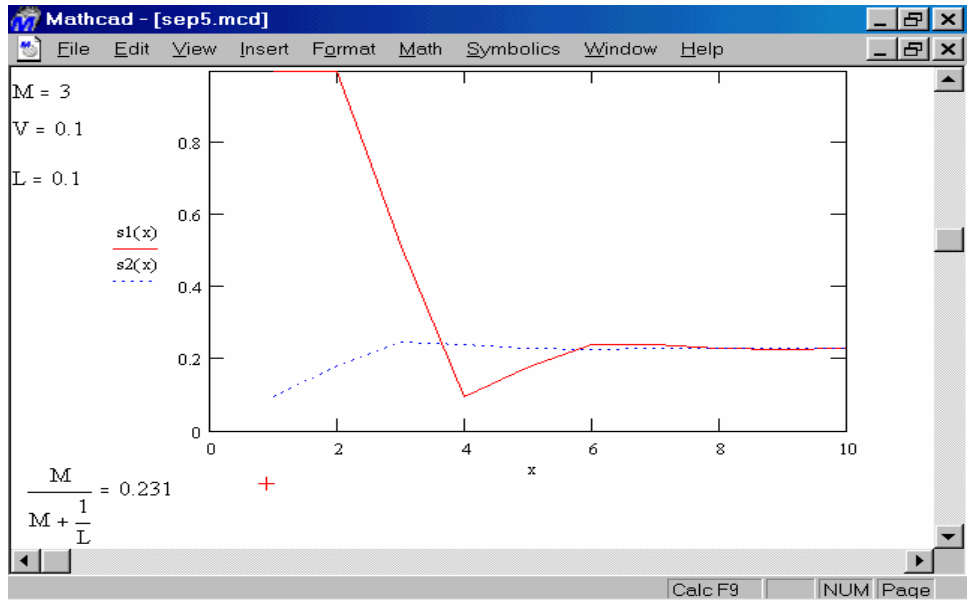

Fig. 7a

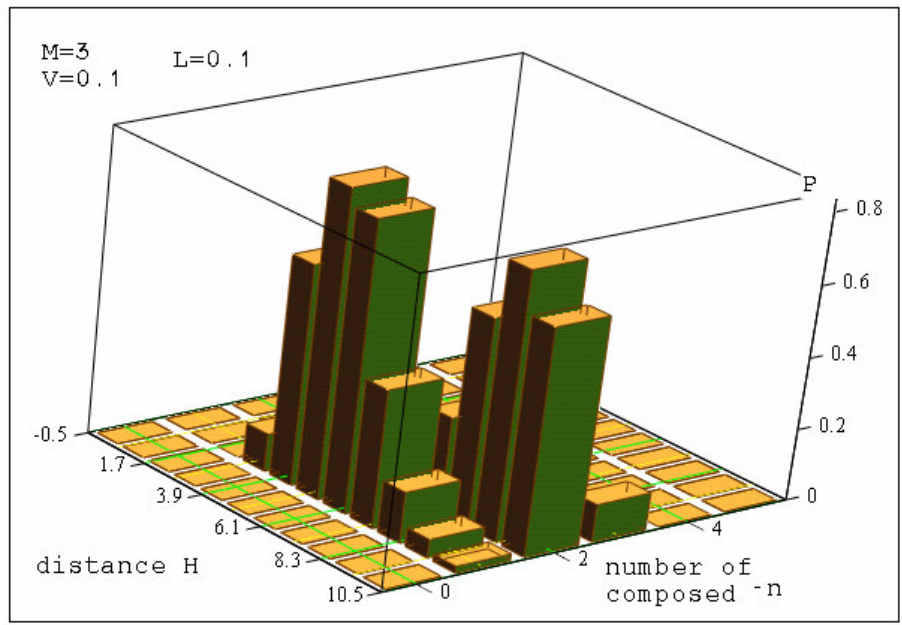

Fig. 7b

Fig. 7. Fig. 7a Probabilities of hit in a package when the previous point was in the beginning of a package s1(x) and in the beginning of an interval between packages s2(x) at rather large distance between packages $M < 1/\lambda$. Variable L in figure corresponds $\lambda$. Variable x in figure corresponds H. Fig. 7b The dependence of size of expression (2. 14) from n and H. is visible what enough to take into account small n

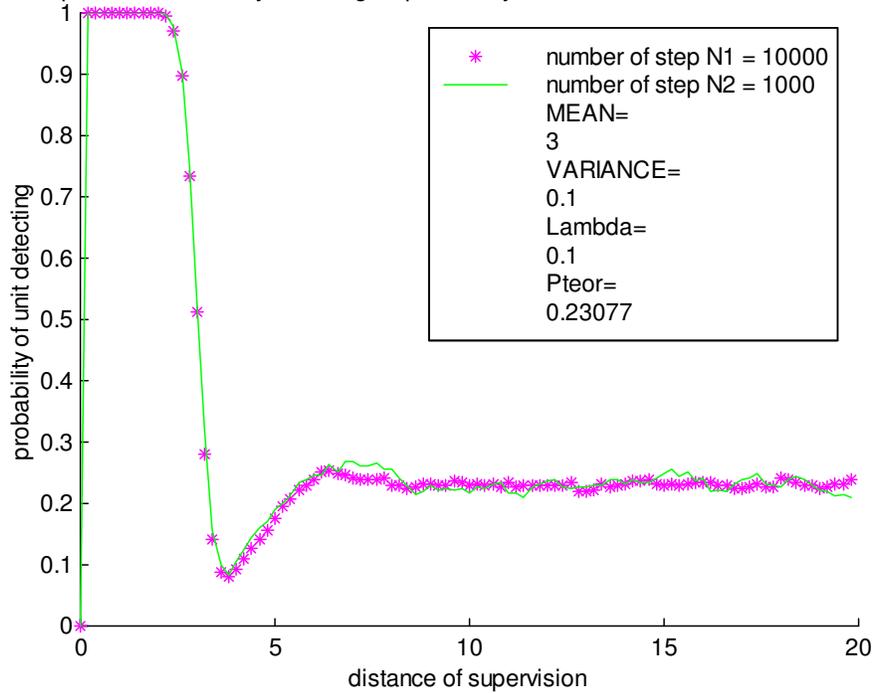

Fig. 8 The Simulated probabilities of hit in a package when the previous point was in the beginning of a package at rather large distance between packages $M < 1/\lambda$ ($\lambda = 0.1$, two dimensions)

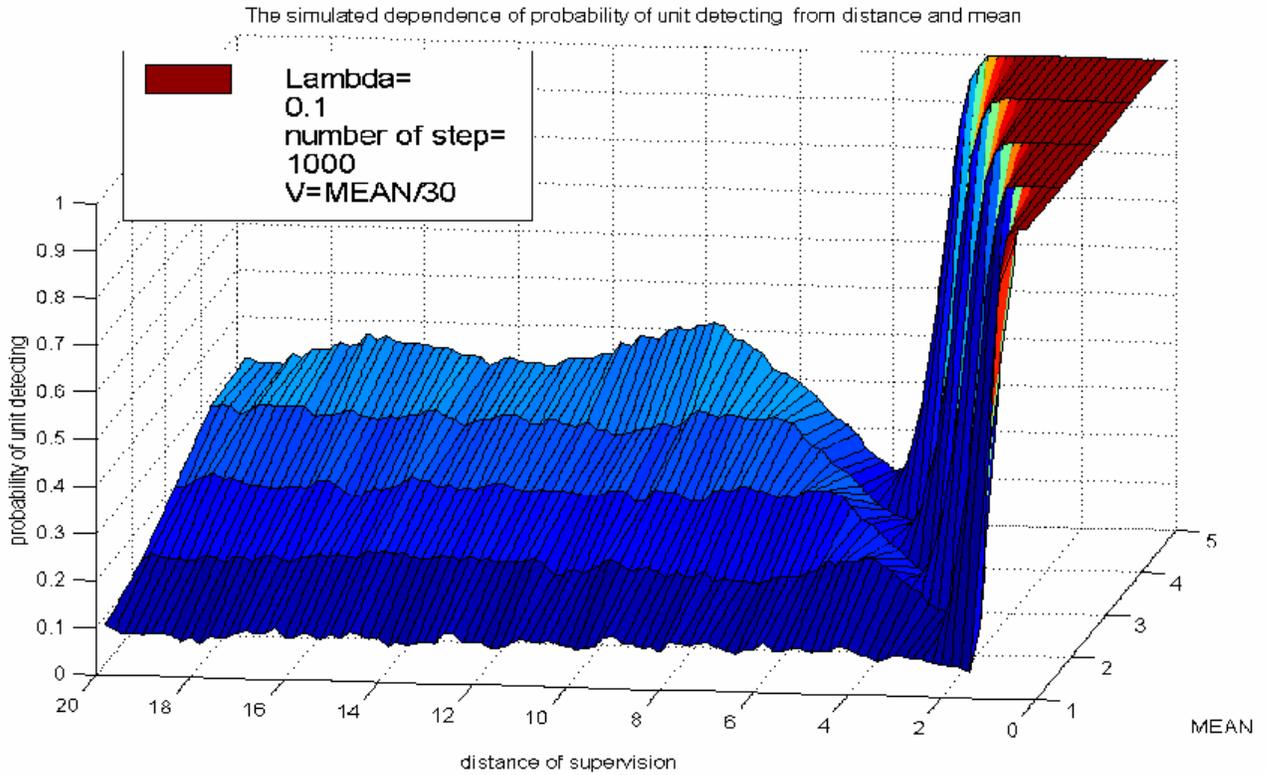

Fig. 9 The Simulated probabilities of hit in a package when the previous point was in the beginning of a package at rather large distance between packages M < 1/$\lambda$ ($\lambda$ = 0.1, three dimensions)

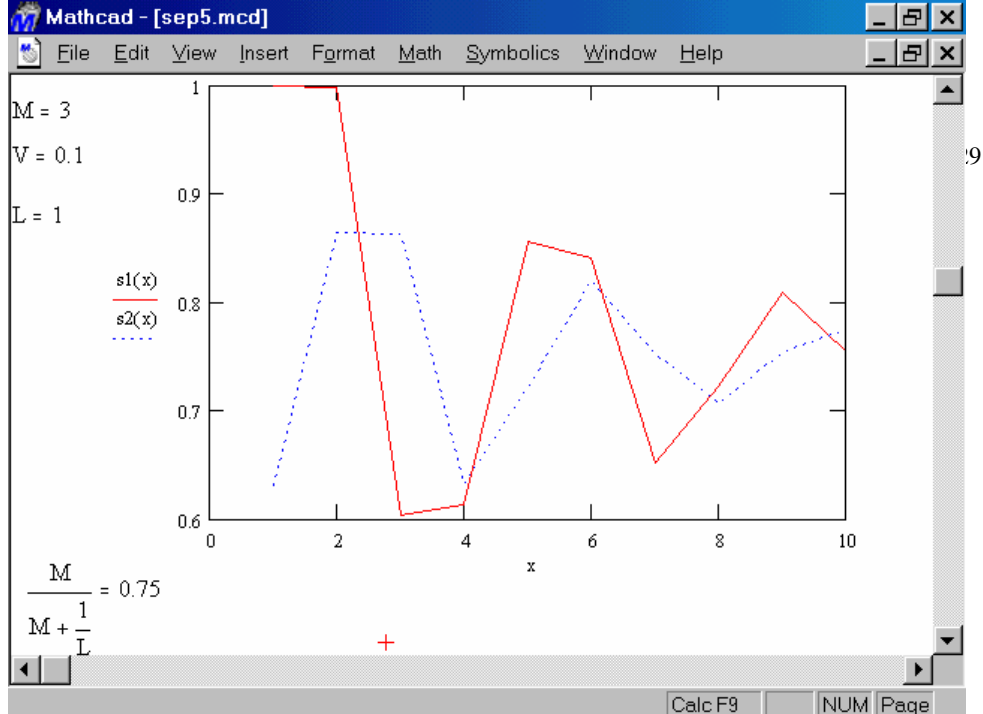

Fig. 10a

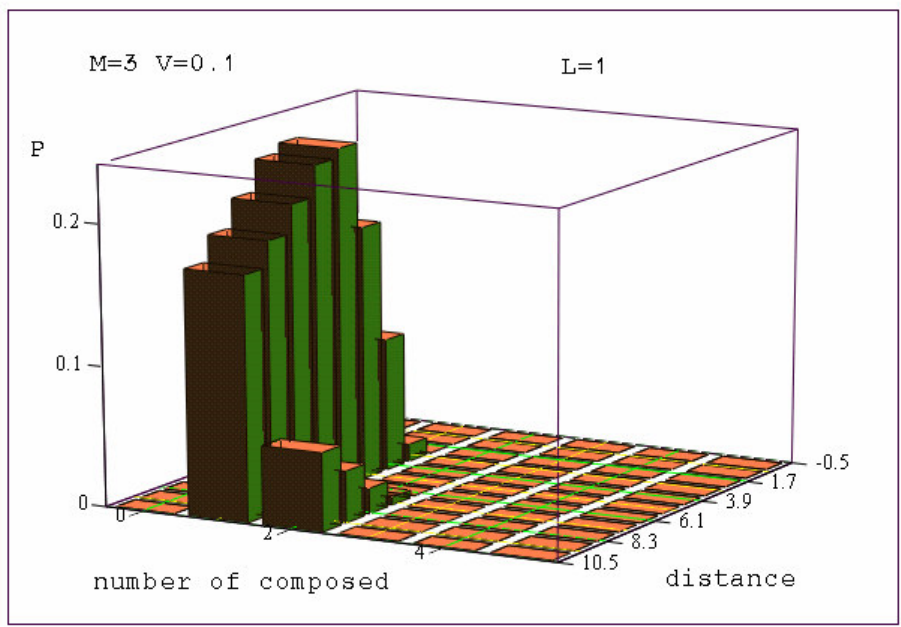

Fig. 10b

Fig. 10 Fig. 10a. Probabilities of hit in a package when the previous point was in the beginning of a package s1(x) and in the beginning of an interval between packages s2(x) at rather small distance between packages $M > 1/\lambda$. Variable L in figure corresponds $\lambda$. Variable x in figure corresponds H. Fig. 10b The dependence of size of expression (2. 14) from n and H. is visible what enough to take into account small n

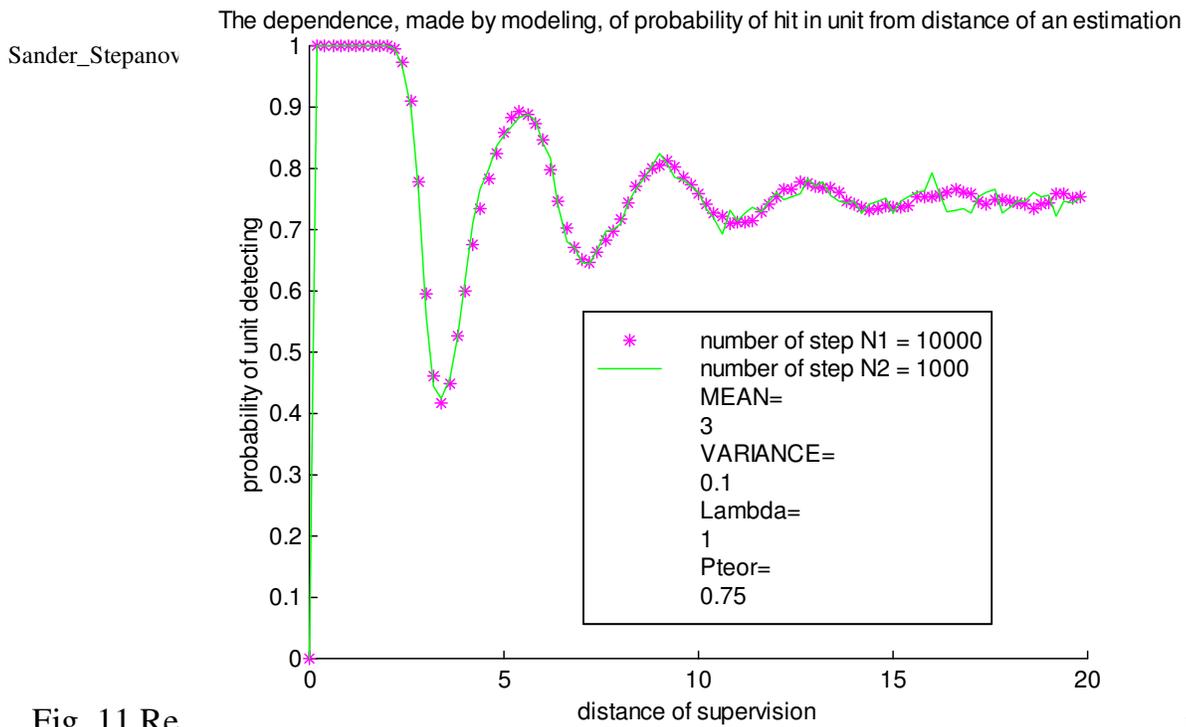

Fig. 11 Re point was in the beginning of a package at rather small distance between packages $M > 1/\lambda$ by means of modeling ($\lambda = 1$, two dimensions).

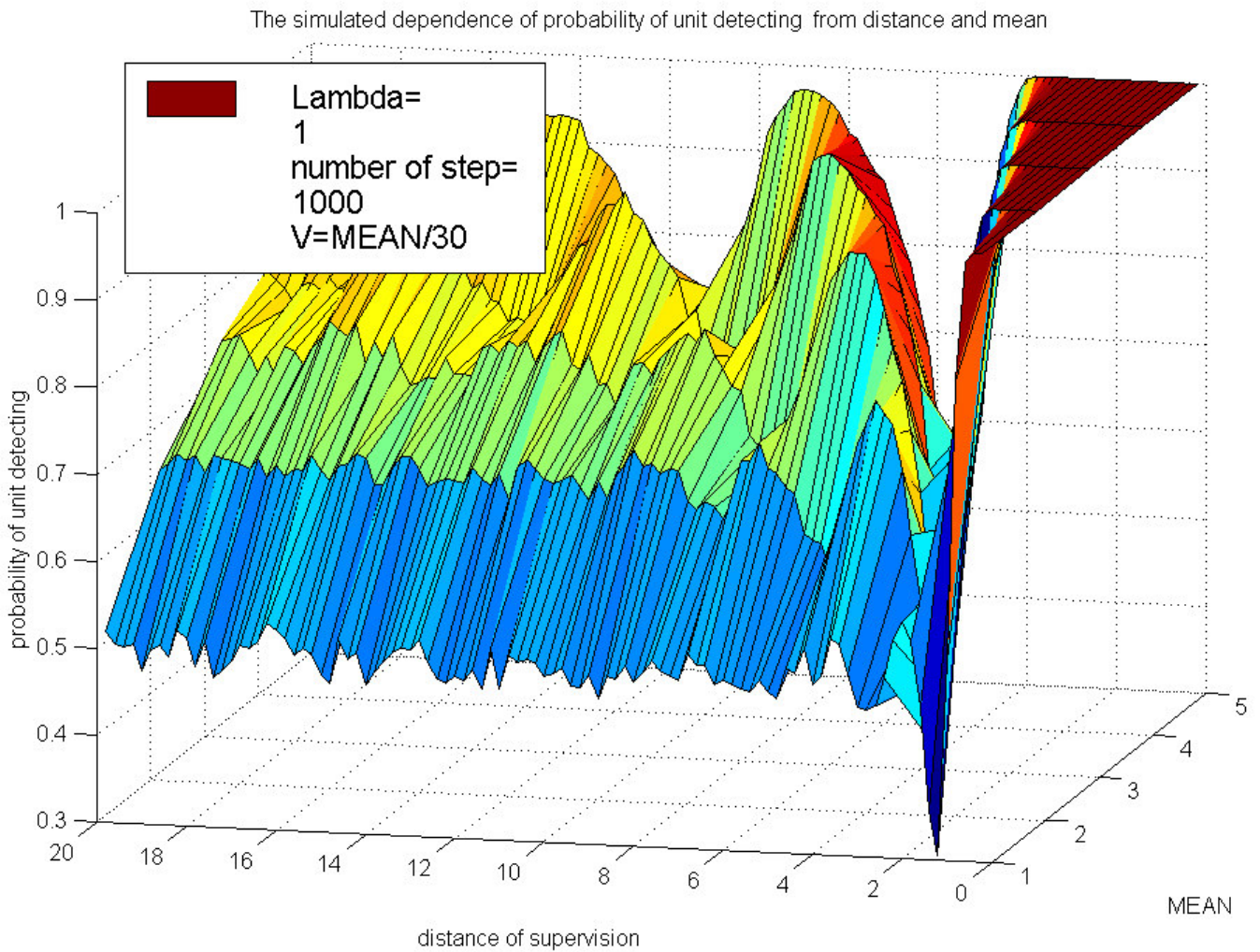

From figures it is visible.

1. Sufficient accuracy of modeling to be made by use 1000 iterations on one point of the diagram.

2. Theoretical results will well be coordinated to experimental results.

$$M := 3 \quad V := 0.1 \quad L := 1 \quad k := 1 \quad N := 10$$

$$x := 1, 2 .. 10$$
$$j := 1, 2 .. N$$

$$f1(x, j) := \left[ \int_0^{x \cdot k} \int_0^y \left(1 - \text{pnorm}(x \cdot k - y, M, \sqrt{V})\right) \cdot \frac{e^{\frac{-(z - M \cdot j)^2}{2 \cdot V \cdot j}}}{\sqrt{V \cdot j \cdot 2 \cdot \pi}} \cdot L \cdot e^{-L \cdot (y - z)} \cdot \frac{(L \cdot (y - z))^{j-1}}{(j - 1)!} \, dz \, dy \right]$$

$$f2(x, j) := \left[ \int_0^{x \cdot k} \int_0^y \left(1 - \text{pnorm}(x \cdot k - y, M, \sqrt{V})\right) \cdot \frac{e^{\frac{-(z - M \cdot j)^2}{2 \cdot V \cdot j}}}{\sqrt{V \cdot j \cdot 2 \cdot \pi}} \cdot L \cdot e^{-L \cdot (y - z)} \cdot \frac{(L \cdot (y - z))^{j}}{(j)!} \, dz \, dy \right]$$

$$f4(x) := 1 - \text{pnorm}(x \cdot k, M, \sqrt{V})$$

$$f3(x) := \left[ \int_0^{x \cdot k} \left(1 - \text{pnorm}(x \cdot k - y, M, \sqrt{V})\right) \cdot L \cdot e^{-L \cdot (y)} \, dy \right] \qquad \frac{M}{M + \frac{1}{L}} = 0.75$$

$$s2(x) := f3(x) + \sum_{j=1}^{N} f2(x, j) \qquad s1(x) := f4(x) + \sum_{j=1}^{N} f1(x, j)$$

Fig. 13 programs of calculation of probability of hit in a package when the previous point was in the beginning s1(x) of a package and in the beginning of an interval between packages s2(x) at rather small distance between packages. Variable L in figure corresponds $\lambda$.

From figures Fig. 7 - Fig 12 it is visible that the various empirical simplifications of expression (2. 11) are possible. For example to calculate expression

$$\left[ QF(\overline{V}_i^{max}, l, H) + \sum_{n=1}^{n=\infty} \int_0^H QF(\overline{V}_i^{max}, l, H - y) bf1(n, y) dy \right] \quad (2.16)$$

for a growing sequence variable H until then while a difference for the next meanings of expression (2. 16) does not become insignificant. If beforehand it is known that the quantity of kinds of lengths of packages does not exceed NLP, it is expedient to apply for f (l) PDF

$$f(l) = \sum_{j=0}^{j=NLP} A_j \delta(l - l_j) \qquad (2.\ 17)$$

were $l_j$ - longs of packets;

$A_j$ - probability of packet with long $l_j$.

Then all integrals on f(l) to be replaced with the sums, that, as it is visible, considerably will simplify calculations.

Step 4. (2.18)

$$P_{post_i}(V_{r_0},...,V_{r_Y},Z_{a_0},..Z_{a_Q}) = \frac{\begin{array}{c}P_{prior}(V_{r_0},...,V_{r_Y},Z_{a_0},..Z_{a_Q}) * P(l_i,V_{r_0},...,V_{r_Y})^{s_i} * \\ * \left[\dfrac{\int_0^\infty f(V_{r_0},...,V_{r_Y},l)l\,dl}{\int_0^\infty f(V_{r_0},...,V_{r_Y},l)l\,dl + \int_0^\infty b(Z_{a_0},..Z_{a_Q},\tau)\tau\,d\tau}\right]^{s_i} \left[1-\dfrac{\int_0^\infty f(V_{r_0},...,V_{r_Y},l)l\,dl}{\int_0^\infty f(V_{r_0},...,V_{r_Y},l)l\,dl + \int_0^\infty b(Z_{a_0},..Z_{a_Q},\tau)\tau\,d\tau}\right]^{1-s_i}\end{array}}{\begin{array}{c}\sum_{r_0=0}^{r_0=m_{0,3}(\bar{V})} \circ\circ\circ \sum_{r_Y=0}^{r_w=m_{w,3}(\bar{V})} \sum_{a_0=0}^{a_0=m_{0,3}(\bar{Z})} \circ\circ\circ \sum_{a_Y=0}^{a_q=m_{q,3}(\bar{Z})} P_{prior}(V_{r_0},...,V_{r_Y},Z_{a_0},..Z_{a_Q}) * P(l_i,V_{r_0},...,V_{r_Y})^{s_i} * \\ * \left[\dfrac{\int_0^\infty f(V_{r_0},...,V_{r_Y},l)l\,dl}{\int_0^\infty f(V_{r_0},...,V_{r_Y},l)l\,dl + \int_0^\infty b(Z_{a_0},..Z_{a_Q},\tau)\tau\,d\tau}\right]^{s_i} \left[1-\dfrac{\int_0^\infty f(V_{r_0},...,V_{r_Y},l)l\,dl}{\int_0^\infty f(V_{r_0},...,V_{r_Y},l)l\,dl + \int_0^\infty b(Z_{a_0},..Z_{a_Q},\tau)\tau\,d\tau}\right]^{1-s_i}\end{array}}$$

for $w = 0,...,Y$; $r = 0,..., m_{w,3}(\bar{V})$; $q = 0,...,Q$; $a = 0,..., m_{q,3}(\bar{V})$;

were $s_i = 1$ if packet is detected, $s_i = 0$ if packet isn't detected by point 1;

$P(l_i, V_{r_0},...,V_{r_Y})^{s_i}$ − likelihood estimation of $l_i$ if $s_i = 1$.

The simplifications described on a step 3 and here will give significant economy in calculations

Step 5    i=i+1

Step 6    $P_{prior_i}(V_{w,r}, Z_{q,a}) = P_{post_{i-1}}(V_{w,r}, Z_{q,a})$ ;

for w=0,…,Y; r=0,…, $m_{w,3}(\overline{V})$; q=0,…,Q; a=0,…, $m_{q,3}(\overline{V})$;

Step 7. Checks on achievement $H_{ind}$ - making independence of samples. That here can be applied with what reasonable method of detection $H_{ind}$. For example use of property of an invariance of probabilities of both kinds of samples. I.e. at use $H_{ind}$ value of expression (1. 5) during estimation should change insignificantly. At the moment of calculation $H_{ind}$ stops of algorithm.

go to Step 2.

Lacks of algorithm 1. Algorithms is deduced not strictly by mathematics, but in the assumption that having calculated $\overline{H_i}$, we with the help it more or less exact knowledge $\overline{H_i}$ can specify knowledge about f(l) and $b(\tau)$, that in turn will allow more precisely to calculate $\overline{H_i}$. Thus, the loop of feedback will be made which will allow quickly and will precisely be adjusted to all monitoring system. In a basis of adjustment of the monitoring system, the loop of feedback lays. As is known, self-excitation is peculiar to a loop of feedback. Therefore, at operation of this algorithm it is necessary to provide the control and prevention of self-excitation. The application of this algorithm together with other less exact but steadier algorithm will be possible by the good conciliatory proposal. For example so: other algorithm solves to a task roughly and after that, the Algorithm 1 improves the decision.

It is necessary to note that the estimation of the traffic can be calculated using vectors of parameters $\overline{V^{max}}$ and $\overline{Z^{max}}$. However at such estimation it is impossible to apply the formulas (1. 21) and (1. 22) for calculation of accuracy of an estimation U. It is possible for calculation of accuracy of an estimation to use posterior probability. However reliability of such characteristic of accuracy is difficult for defining theoretically. In experiments with modeling of estimation U by means of posterior the probability has shown good serviceability.

### 3. 2. Obtaining independence by calculation of conditional probabilities

More simple way of maintenance of independence is the direct calculation of dependence of samples. If will appear that samples are dependent then it will be necessary to calculate distances between points of the control by Search Methods of optimization for function of one variable.

The version of the decision with a direction on a practicality further is stated.

Half-philosophical questions as for example. Where the independence begins? Or where the independence comes to an end? To be considered, as it and was earlier, will not be. Though, at the end as it always happens in practice, during transition from the theory to practice it is necessary to give on them the answers. However, this on the one hand difficult and on the other hand simple work will be transferred on then. As well as in stated earlier, on these difficult questions requiring empirical decision (so-called engineering decisions) will be paid attention and their empirical character will be emphasized.

The independence of samples can be reached by calculation of unconditional probabilities of detection and not detection of a package with their subsequent comparison with conditional and joint probabilities. Let us designate event of hit in a package through 0, event not hits in a package through 1, then for independent samples (P (x/y) - means probability to pull out x if before x have pulled out y;

P (x∩y) - means probability to pull out x and y in the next points)

$$P(1) \approx P(1/0) \approx P(1/1) \qquad P(0) \approx P(0/1) \approx P(0/0) \quad (2.\ 17)$$

$$P(1 \cap 1) \approx P(1)P(1); \quad P(1 \cap 0) \approx P(1)P(0); \qquad (2.\ 18)$$

$$P(0 \cap 0) \approx P(0)P(0); \quad P(0 \cap 1) \approx P(0)P(1);$$

The accuracy of calculations can be supervised using expressions (1. 21), (1. 22). Thus, there is a danger of wrong work expressions (2. 17, 2.18) because of dependence of samples. If it to be found out then it is necessary to develop algorithms of the control of accuracy of calculation of probabilities taking place in expressions (2. 17), (2. 18).

We shall consider on an example of a detail of calculation (2. 17), (2. 18). On Fig. 14 the typical situation on an output of a source is shown. For bad extraction of samples ( the samples are dependent) of a point of the control are designated through "-", for good extraction of samples (the samples are independent) of a point of the control are designated through "+". Let's designate their "extraction –" and "extraction +".

On Fig. 15 and Fig. 16 is shown what picture sees "extraction –" and "extraction +". Accounts for "extraction – " ("yes" and "not" designate conformity or not conformity to theoretical expectations).

$$P(1) = 9/15 \neq P(1/0) = 1/10 \neq P(1/1) = 8/9;\ yes$$
$$P(0) = 6/15 \neq P(0/1) = 1/9 \neq P(0/0) = 4/6;\ yes$$
$$P(1 \cap 1) = 8/14 \neq P(1)P(1) = 0.36,\quad P(1 \cap 0) = 1/14 \neq P(1)P(0) = 0.24;\ yes$$
$$P(0 \cap 0) = 4/14 \approx P(0)P(0) = 0.16;\ not$$
$$P(0 \cap 1) = 1/14 = 0.07 \neq P(0)P(1) = 0.24;\ yes$$

Accounts for "extraction +".

$$P(1) = 5/15 = 0.3 \approx P(1/0) = 3/10 = 0.3 \approx P(1/1) = 1/5 = 0.2;\ yes$$
$$P(0) = 10/15 = 0.7 \approx P(0/1) = 4/5 = 0.8 \approx P(0/0) = 6/10 = 0.6;\ yes$$
$$P(1 \cap 1) = 1/14 = 0.07 \approx P(1)P(1);\quad P(1 \cap 0) = 3/14 = 0.02 \approx P(1)P(0);\ yes$$
$$P(0 \cap 0) = 6/14 = 0.4 \approx P(0)P(0);\quad P(0 \cap 1) = 3/14 = 0.2 \approx P(0)P(1);\ yes$$

As it is visible from an example for revealing dependence it is necessary to take into account all possible combinations "0" and "1".

## The conclusion.

1. The offered estimation of the traffic calculates the characteristics of accuracy of an estimation of the traffic before an estimation and during process of an estimation. A little of time of an estimation is achieved by: a) independent samples get that provides a plenty of the information about the researched traffic; b) intervals of time between the next samples are minimized with preservation of their independence that results in minimization of common time of supervision.

2. The reliability of the achieved results is provided theoretically and experimental modeling.

3. The advanced theory can be used for an estimation as already of transferred traffic and for a prediction of the future traffic.

3.        Offered approach can be applied in…

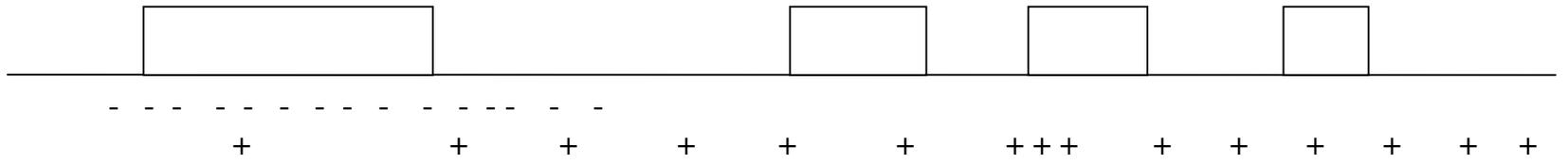

Fig. 14

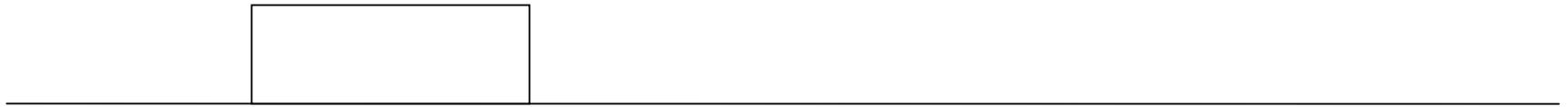

Fig. 15

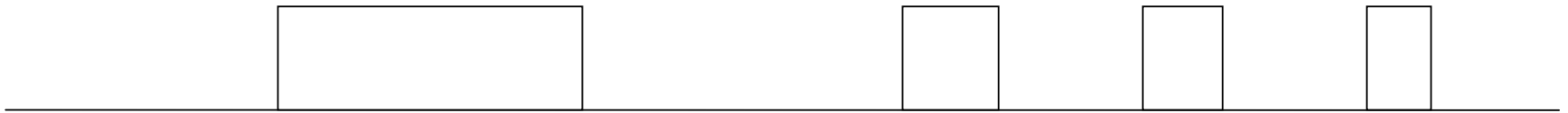

Fig. 16